\documentclass{elsart}

\usepackage{amsmath}
\usepackage{amsfonts}
\usepackage{mathrsfs,cite}

\newcommand{\beqs}{\begin{equation*}}
\newcommand{\beq}{\begin{equation}}

\newcommand{\eeqs}{\end{equation*}}
\newcommand{\eeq}{\end{equation}}

\newcommand{\beqas}{\begin{eqnarray*}}
\newcommand{\beqa}{\begin{eqnarray}}

\newcommand{\eeqas}{\end{eqnarray*}}
\newcommand{\eeqa}{\end{eqnarray}}

\newcommand{\eq}[2]{\begin{equation} #1 \label{#2} \end{equation}}

\newcommand{\ga}{\gamma}

\newcommand{\om}{\omega}

\newcommand{\blist}{\begin{itemize}}

\newcommand{\elist}{\end{itemize}}

\providecommand{\href}[2]{#2}

\DeclareFontFamily{OT1}{rsfs}{}
\DeclareFontShape{OT1}{rsfs}{m}{n}{ <-7> rsfs5 <7-10> rsfs7 <10->rsfs10}{} 
\DeclareMathAlphabet{\mycal}{OT1}{rsfs}{m}{n}

\DeclareMathOperator{\extdm}{d}
\newcommand{\extd}{\extdm \!}

\newcommand\MM{\mathcal{M}}
\newcommand\dM{\partial \MM}

\newcommand{\Lag}{{\cal L}}

\journal{Physics Letters B}

\begin{document}

\begin{frontmatter}



\hfill MIT--CTP 4011,
%
TUW 09--03,
%
YITP--SB--09--04

\title{Holographic counterterms from local supersymmetry without boundary conditions}

\author[DG]{Daniel Grumiller}
\ead{grumil@hep.itp.tuwien.ac.at}
and
\author[PvN]{Peter van Nieuwenhuizen}
\ead{vannieu@insti.physics.sunysb.edu}

\address[DG]{Center for Theoretical Physics, Massachusetts Institute of Technology,\\ 77 Massachusetts Ave, Cambridge, MA 02139, USA \\ and \\ Institute for Theoretical Physics, Vienna University of Technology,\\ Wiedner Hauptstr.~8--10/136, Vienna, A-1040, Austria}

\address[PvN]{C.N. Yang Institute for Theoretical Physics, Math 6-118, Stony Brook University,\\ Stony Brook, NY 11794-3840, USA}

\begin{abstract}
We show in some lower-dimensional supergravity models that the holographic counterterms which are needed in the AdS/CFT correspondence to make the theory finite, coincide with the counterterms that are needed to make the action supersymmetric without imposing any boundary conditions on the fields. 
\end{abstract}


\end{frontmatter}

\section{Introduction}

Holographic renormalization is the by now well-known procedure of subtracting boundary counterterms from the action in order to render the variational principle well-defined. As an additional benefit the boundary-stress tensor usually becomes finite by this procedure, which provided the original motivation for holographic renormalization \cite{Balasubramanian:1999re,Emparan:1999pm,Myers:1999psa,deHaro:2000xn,Bianchi:2001kw,Henningson:1998gx,
Skenderis:2002wp,Papadimitriou:2004ap,Mann:1999pc}. 
The full action
\eq{
I=I_{\textrm{\tiny bulk}} + I_{\textrm{\tiny GHY}} - I_{\textrm{\tiny counter}}
}{eq:pvn1}
consists of three parts: a bulk action $I_{\textrm{\tiny bulk}}$ that generates the equations of motion (EOMs), a Gibbons--Hawking--York (GHY) boundary term $ I_{\textrm{\tiny GHY}}$ that leads to a Dirichlet boundary value problem, and the holographic counterterm $I_{\textrm{\tiny counter}}$. The latter guarantees that the first variation of the full action vanishes for all variations that preserve the boundary conditions for the fields:
\eq{
\delta I = 0
}{eq:pvn2}
While there are different techniques --- for instance, the Hamilton--Jacobi method \cite{deBoer:1999xf, Martelli:2002sp, Larsen:2004kf, Batrachenko:2004fd} --- to implement the procedure of holographic renormalization, all of them have one feature in common: they require the specification of precise boundary conditions for the fields.

In applications this procedure works well, but conceptually it is not entirely satisfying, for the following logical cycle arises: in order to determine the appropriate boundary conditions for the fields one should know already the behavior of classical solutions near the boundary. This requires knowledge of the EOMs, which are derived from the action. But in order to {\em consistently} derive the EOMs by virtue of a variational principle from the action, one has to know the action including all boundary terms. If the derivation of the latter appeals to appropriate boundary conditions we are back to square one.

In this note we show that supersymmetry (SUSY) can break the logical cycle. Namely, we apply the main credo of \cite{Belyaev:2005rt,Belyaev:2007bg,Belyaev:2008xk}: an action should be SUSY-invariant, even in the presence of boundaries, {\em without imposing any boundary conditions on the fields}. In other words, off-shell there are no boundary conditions needed for maintaining SUSY. 
 Since many of the theories where the issue of holographic renormalization arises are supergravity (SUGRA) theories, the requirement of SUSY is pertinent. Moreover, the main credo jibes well with our desire to avoid imposing boundary conditions on the fields in order to evade the logical cycle. However, at this point it is not at all clear if SUSY has anything to say about holographic renormalization. The main purpose of our note is to exhibit that this is indeed the case.

For sake of specificity we restrict ourselves initially to a study of 3-dimensional SUGRA theories. The simplest example is pure SUGRA with negative cosmological constant, whose bosonic version of the action $I$ is given by the bulk action
\eq{
I_{\textrm{\tiny bulk}} = \frac12 \int_\MM \!\!\!\extd^3x\sqrt{-g}\,\Big(R-\frac{2}{\ell^2}\Big)
}{eq:pvn0}
and the boundary action $I_b=I_{\textrm{\tiny GHY}}-I_{\textrm{\tiny counter}}$, with \cite{Balasubramanian:1999re,Emparan:1999pm,Myers:1999psa,deHaro:2000xn,Skenderis:2002wp}
\begin{subequations}
\label{eq:boundary}
\begin{align}
I_{\textrm{\tiny GHY}} &= \int_{\dM}\!\!\!\!\!\!\extd^2x\sqrt{-h}\, K \label{eq:GHY} \\
I_{\textrm{\tiny counter}} &= \int_{\dM}\!\!\!\!\!\!\extd^2x\sqrt{-h}\, \frac{1}{\ell} \label{eq:counter} 
\end{align}
\end{subequations}
Here $h$ is the determinant of the induced metric at the boundary $\dM$ of the spacetime $\MM$ and $K$ is the trace of extrinsic curvature.
Our goal is to derive the boundary action \eqref{eq:boundary} from the knowledge of the bulk action \eqref{eq:pvn0} by imposing local SUSY. It is non-trivial and interesting that this is possible.

To set up the stage we summarize in section \ref{se:2} the results of \cite{Belyaev:2007bg}, which lead to the GHY boundary term \eqref{eq:GHY}. In section \ref{se:3} we show that SUSY without boundary conditions automatically leads to the correct holographic counterterm \eqref{eq:counter}. To investigate whether our conclusions apply also to other cases we consider generic 2-dimensional dilaton SUGRA in section \ref{se:4}. We find again that imposing SUSY without boundary conditions establishes the correct holographic counterterm. 

\section{Review of SUGRA without boundary conditions}\label{se:2}

We review now briefly the results of \cite{Belyaev:2007bg}, whose conventions we adopt: 
Our Ricci-scalar is positive for AdS. We set $8\pi G_N=1$ and use signature $(-,+,+)$. Upper case indices refer to the bulk theory and lower case indices to the boundary theory. Indices from the beginning of the alphabet ($A,B,\dots$ and $a,b,\dots$) refer to an anholonomic frame (``flat indices'') and indices from the middle of the alphabet ($M,N,\dots$ and $m,n,\dots$) refer to a holonomic frame (``curved indices''). The boundary $\dM$ is a surface of constant $x^3$, located at $x^3=0$ (in the bulk $x^3>0$). The Lorenz gauge $e_a{}^3=0$ is imposed (and thus $e_m{}^{\hat 3}=0$). Defining $\epsilon_\pm=\frac12 (1\pm\ga^{\hat 3})\epsilon$ where $\ga^{\hat 3}$ is constant, the unbroken half of SUSY is generated by $\epsilon_+$. When considering SUSY transformations in this paper we always mean the ``modified $\epsilon_+$ SUSY'' of \cite{Belyaev:2007bg}, a specific linear combination of SUSY and compensating Lorentz-transformations that preserves the Lorenz gauge condition $e_m{}^{\hat 3}=0$.  The quantity $\widehat\om$ and related hatted curvature quantities such as extrinsic curvature $\widehat K$ or gravitino field strength $\widehat\psi_{MN}=\widehat D_M\psi_N-\widehat D_N\psi_M$ always refer to supercovariant objects. 

We are interested in constructing SUGRA actions of the form
\eq{
I=\int_{\MM}\!\!\!\extd^3x\, \Lag_F - \int_{\dM}\!\!\!\!\!\!\extd^2x\, \Lag_A
}{eq:pvn4}
and require that half of SUSY and diffeomorphisms along the boundary are preserved:
\eq{
\delta_\xi I = 0 \qquad \delta^\prime_{\epsilon_+} I = 0
}{eq:pvn7}
Here $\xi$ refers to 2-dimensional diffeomorphisms within the boundary and $\delta^\prime_{\epsilon_+}$ refers to the unbroken modified SUSY transformations. It is crucial for the credo stated in the introduction that \eqref{eq:pvn7} holds without imposing boundary conditions on the fields. If this is the case then we call an action \eqref{eq:pvn4} with the property \eqref{eq:pvn7} SUSY-invariant.\footnote{By ``SUSY-invariant'' in this note we always mean ``locally SUSY-invariant with respect to modified $\epsilon_+$ SUSY and without imposing any boundary conditions on the (off-shell) fields''.}
Suppose that in addition to the SUGRA multiplet
\eq{
\big(e_M{}^A,\quad\psi_M,\quad S\big)
}{eq:pvn12}
we have a (composite or fundamental) scalar multiplet
\eq{
\big(A,\quad\chi,\quad F\big)
}{eq:pvn8}
The main result of \cite{Belyaev:2007bg} is that the SUSY-invariant action is given by
\eq{
I=\int_{\MM}\!\!\!\extd^3x \underbrace{e_3\, \Big(F+\frac12\bar\psi_M\gamma^M\chi+\frac14A\bar\psi_M\ga^{MN}\psi_N+AS\Big)}_{{\cal L}_F} - \int_{\dM}\!\!\!\!\!\!\extd^2x \underbrace{e_2\,A}_{{\cal L}_A}
}{eq:pvn9}
The bulk action in $3+1$ dimensions was obtained in \cite{Ferrara:1978wj} 
while the $2+1$ and $1+1$ cases were obtained in \cite{Uematsu:1984zy}. 
In the case of pure SUGRA the relevant multiplet realizing \eqref{eq:pvn8} is the the scalar curvature multiplet
\eq{
\big(S,\quad\frac12\gamma^{MN}\widehat\psi_{MN}-\frac12 \gamma^M\psi_M S,\quad\frac12 R(\widehat\om) -\frac12 \bar\psi^M\gamma^N\widehat\psi_{MN}+\frac14S\bar\psi^M\psi_M-\frac34 S^2\big)
}{eq:pvn10}

The SUSY-invariant action \eqref{eq:pvn9} entails an ambiguity. Namely, consider a co-dimension 1 spinor multiplet
$(\chi^\prime,\; A^\prime)$
whose highest component contributes to the boundary action. This implies a shift of the boundary Lagrange-density $\Lag_A\to \Lag_A+\Lag_{A^\prime}$. Such a shift is possible because $\Lag_{A^\prime}$ is SUSY-invariant by itself. As demonstrated in \cite{Belyaev:2007bg} the ambiguity of adding a SUSY-invariant boundary term can be fixed uniquely for pure SUGRA: without an appropriate co-dimension 1 multiplet the boundary action would contain the auxiliary field $S$ linearly, which in turn would imply the (unphysical) boundary EOM $e_2=0$. The boundary term linear in $S$ can be cancelled with a boundary term constructed from the extrinsic curvature multiplet
$(\ga^a\psi_{a-},\;\widehat K+S)$.
The result for the SUSY invariant action of 3-dimensional pure SUGRA is
\eq{
I_{\textrm{\tiny SUGRA}}=\frac12 \int_{\MM}\!\!\!\extd^3x e_3\,\Big(R(\widehat\om)+\bar\psi_M\ga^{MNK}\widehat D_N\psi_K+\frac12 S^2\Big) 
+\int_{\dM}\!\!\!\!\!\!\extd^2x e_2 \,\Big(\widehat K +\frac12\bar\psi_{a+}\ga^a\ga^b\psi_{b-}\Big)
}{eq:pvn15}
Setting the gravitino $\psi_M$ to zero and eliminating the auxiliary field $S$ by means of its EOM leads to the Einstein--Hilbert action \eqref{eq:pvn0} with the GHY boundary term \eqref{eq:GHY}. The result \eqref{eq:pvn15} was derived without imposing any boundary conditions on the fields \cite{Belyaev:2007bg}. There is no counterterm of the form \eqref{eq:counter} in this example because we are not yet in AdS space.

\section{Holographic counterterms from SUGRA without boundary conditions}\label{se:3}

The issue of holographic renormalization typically arises for theories where the EOMs lead to solutions for the metric that asymptote to AdS \cite{Skenderis:2002wp}. We are therefore led to consider 3-dimensional SUGRA theories that allow for asymptotically AdS solutions. The simplest one is obtained from pure SUGRA by adding a cosmological constant multiplet
\eq{
\big(\frac{1}{\ell},\quad 0,\quad 0\big)
}{eq:pvn17}
where $\ell$ is the AdS radius. Inserting $A=\frac1\ell$, $\chi=0$, $F=0$ into the result for the SUSY-invariant action \eqref{eq:pvn9} yields the SUSY-invariant cosmological constant action
\eq{
I_\Lambda=\int_{\MM}\!\!\!\extd^3x e_3\,\Big(\frac{1}{4\ell}\bar\psi_M\ga^{MN}\psi_N+\frac1\ell S\Big)-\int_{\dM}\!\!\!\!\!\!\extd^2x e_2\,\frac{1}{\ell}
}{eq:pvn18}
The total SUSY-invariant action $I_{\Lambda\textrm{\tiny SUGRA}}$ is obtained by adding the cosmological constant action \eqref{eq:pvn18} to the SUGRA action \eqref{eq:pvn15}. The result is\footnote{Using a similar philosophy as in the present work, Luckock and Moss derived the action \eqref{eq:pvn19} to order fermion-squared \cite{Luckock:1989jr} (see also \cite{Moss:2004ck}), which happens to be the complete result as shown in the present work.} 
\begin{align}
I_{\Lambda\textrm{\tiny SUGRA}} &= \frac12 \int_{\MM}\!\!\!\extd^3x e_3\,\Big(R(\widehat\om)+\bar\psi_M\ga^{MNK}\widehat D_N\psi_K+\frac{1}{2\ell}\bar\psi_M\ga^{MN}\psi_N+\frac12 S^2+\frac2\ell S\Big) \nonumber \\
&\quad\,+\int_{\dM}\!\!\!\!\!\!\extd^2x e_2 \,\Big(\widehat K -\frac{1}{\ell} +\frac12\bar\psi_{a+}\ga^a\ga^b\psi_{b-}\Big) 
\label{eq:pvn19}
\end{align}
Setting the gravitino $\psi_M$ to zero and eliminating the auxiliary field $S$ by means of its EOM leads to the bosonic action
\eq{
I_{\Lambda\textrm{\tiny EH}}=\frac12 \int_{\MM}\!\!\!\extd^3x \sqrt{-g}\,\Big(R-\frac{2}{\ell^2}\Big)+\int_{\dM}\!\!\!\!\!\!\extd^2x \sqrt{-h} \,\Big(K-\frac{1}{\ell}\Big)
}{eq:pvn22}
This is the cosmological Einstein--Hilbert action with the GHY boundary term \eqref{eq:GHY} and the holographic counterterm \eqref{eq:counter}.\footnote{Of course, one ambiguity always remains: we can add an arbitrary (finite or infinite) constant to the action, like the logarithmic subtraction linked with the Weyl anomaly \cite{Henningson:1998gx}. This ambiguity cannot be fixed at the level of the action, but only upon appealing to specific solutions of the EOM, e.g.~by demanding that the free energy of the ground state solution vanishes.} Thus, we have reached our goal to derive the result for the holographic counterterm from requiring SUSY-invariance of the action. We achieved this without imposing any boundary conditions on the fields.

\section{Two dimensional dilaton SUGRA}\label{se:4}

So far we have treated SUGRA in three spacetime dimensions. To investigate whether our conclusions apply also to other cases we consider here 2-dimensional dilaton SUGRA.
Dilaton SUGRA in two dimensions was introduced by Park and Strominger \cite{Park:1993sd}, based upon the work by Howe \cite{Howe:1979ia}. It was studied in detail e.g.~in \cite{Izquierdo:1998hg,Ertl:2000si,Grumiller:2002nm,Bergamin:2002ju,Bergamin:2003mh}.  

We need the following multiplets. The 2-dimensional curvature multiplet \cite{Uematsu:1984zy}
\eq{
\big(S,\quad\underbrace{\frac12\gamma^{MN}\widehat\psi_{MN}-\frac12 \gamma^M\psi_M S}_{:=\zeta},\quad\frac12 R(\widehat\om)-\frac12 \bar\psi^M\gamma^N\widehat\psi_{MN}+\frac14S\bar\psi^M\psi_M-S^2\big)
}{eq:pvn40}
the dilaton multiplet
\eq{
\big(X,\quad\chi,\quad F\big)
}{eq:pvn29}
and the pre-potential multiplet
\eq{
\big(u(X),\quad u^\prime(X)\chi,\quad u^\prime(X)F-\frac12 u''(X) \bar\chi\chi\big)
}{eq:pvn38}
The bosonic bulk action without auxiliary fields is of the form
\eq{
I_{\textrm{\tiny DG}}^{\textrm{\tiny bulk}}=\frac12 \int_{\MM}\!\!\!\extd^2x\sqrt{-g}\,\Big(XR - 2u(X)u^\prime(X)\Big)
}{eq:pvn25}
In \cite{Grumiller:2007ju} the full boundary action $I_b=I_{\textrm{\tiny DGHY}}-I_{\textrm{\tiny Dcounter}}$ was derived using the Hamilton--Jacobi method of holographic renormalization:
\begin{subequations}
\label{eq:pvnb}
\begin{align}
I_{\textrm{\tiny DGHY}} &=\int_{\dM}\!\!\!\!\!\!\extd^2x \sqrt{-h} \, XK \label{eq:pvn25a} \\
I_{\textrm{\tiny Dcounter}} &=\int_{\dM}\!\!\!\!\!\!\extd^2x \sqrt{-h} \, u(X) \label{eq:pvn25b}
\end{align}
\end{subequations}
Our goal to derive the boundary terms \eqref{eq:pvnb} from SUSY-invariance. 

The product of the curvature multiplet \eqref{eq:pvn40} and the dilaton multiplet \eqref{eq:pvn29} leads to
\eq{
\big(SX,\quad S\chi+X\zeta,\quad\frac12 XR(\widehat\om)-XS^2+SF-\bar\chi\zeta-\frac12 X\bar\psi^M\ga^N\widehat\psi_{MN}+\frac14 XS\bar\psi^M\psi_M\big)
}{eq:pvn39} 
For a scalar multiplet $(\tilde A,\;\tilde\chi,\;\tilde F)$ the 2-dimensional version of the SUSY-invariant action \eqref{eq:pvn9} is given by
\eq{
I=\int_{\MM}\!\!\!\extd^2 x e_2\, \Big(\tilde F+\frac12\,\bar\psi_M\ga^M\tilde \chi+\frac14\,\tilde A\bar\psi_M\ga^{MN}\psi_N+\tilde A S\Big) - \int_{\dM}\!\!\!\!\!\!\extd x e_1\, \tilde A
}{eq:pvn137}
We plug now the multiplet \eqref{eq:pvn39} into the 2-dimensional SUSY-invariant action \eqref{eq:pvn137}, then we do the same with the pre-potential multiplet \eqref{eq:pvn38} and add both contributions. This procedure yields an action $\tilde I_{\textrm{\tiny DSG}}$ that contains a boundary term linear in the auxiliary field. Such a term is problematic, because elimination of the auxiliary field $S$ implies an unphysical boundary EOM, $e_1 X=0$.\footnote{Setting $e_1=0$ leads to a degenerate boundary metric. Setting $X=0$ eliminates the space of solutions and implies infinite gravitational coupling at the boundary.} In order to cancel the offending term we employ the co-dimension 1 multiplet
\eq{
\big(X\ga^a\psi_{a-},\quad X(\widehat K+S)-\chi_-\ga^a\psi_{a-}\big)
}{eq:pvn42.2} 
and add the corresponding SUSY-invariant boundary action to the action $\tilde I_{\textrm{\tiny DSG}}$. This obtains uniquely the SUSY invariant dilaton SUGRA action
\begin{multline}
I_{\textrm{\tiny DSG}} = \frac12\int_{\MM}\!\!\!\extd^2x e_2 \,\Big(XR(\widehat\om)+2SF+2u(X)S+2u^\prime(X)F-\bar\chi\ga^{MN}\widehat\psi_{MN} \\
+\frac12 u(X)\bar\psi_M\ga^{MN}\psi_N+u^\prime(X)\bar\psi^M\ga^M\chi-u''(X)\bar\chi\chi\Big) \\
+\int_{\dM}\!\!\!\!\!\!\extd^2x e_1\, \Big(X\widehat K-u(X)+\frac12 X\bar\psi_{a+}\ga^a\ga^b\psi_{b-}-\chi_-\ga^a\psi_{a-}\Big)
\label{eq:pvn41a}
\end{multline}
The bulk part of the action \eqref{eq:pvn41a} up to notational changes and integrating out auxiliary fields coincides with the actions used in \cite{Park:1993sd,Izquierdo:1998hg,Ertl:2000si,Grumiller:2002nm,Bergamin:2002ju,Bergamin:2003mh}. The boundary part of the action \eqref{eq:pvn41a} required to maintain SUSY-invariance is a new result. 

Integrating out the auxiliary field density $e_2 S$ leads to a functional delta-function whose argument implies the constraint
$F=-u(X)$
for the dilaton auxiliary field $F$ upon integrating out the latter. No Jacobians arise from these integrations. Setting all spinors to zero and integrating out $S$ and $F$, the action \eqref{eq:pvn41a} reduces to
\eq{
I_{\textrm{\tiny DG}} = \frac12\int_{\MM}\!\!\!\extd^2x \sqrt{-g} \,\Big(XR-2u(X) u(X)^\prime\Big)+\int_{\dM}\!\!\!\!\!\!\extd^2x \sqrt{-h} \,\Big(XK-u(X)\Big)
}{eq:pvn44}
Comparison of the bulk action in \eqref{eq:pvn44} with the bulk action \eqref{eq:pvn25} shows that they coincide. Comparison of the boundary action in \eqref{eq:pvn44} with the boundary action \eqref{eq:pvnb} establishes again the remarkable result that SUSY-invariance automatically leads to the GHY boundary term \eqref{eq:pvn25a} and to the holographic counterterm \eqref{eq:pvn25b}. 
The result \eqref{eq:pvn44} was derived without imposing any boundary conditions on the fields. 

For simplicity we have neglected a kinetic term for the dilaton. It can be introduced through a dilaton-dependent Weyl rescaling, see for instance section 5 in \cite{Bergamin:2003am} for details. It can be checked easily that our procedure leading to \eqref{eq:pvn44} generalizes to models containing a kinetic term for the dilaton field. In this way we recover (up to notational changes) Eq.~(7.1) of \cite{Grumiller:2007ju}.\footnote{The relation between pre-potential $u(X)$ and various functions of the dilaton is as follows: $e^{-Q(X)}w(X)=u^2(X)$. For the case of interest $Q(X)=U(X)=0$ we have the simple relations $V(X)=-u(X)u^\prime(X)$ and $w(X)=u^2(X)$.}

\section{Conclusions}

We have demonstrated in this note that imposing SUSY without boundary conditions on the fields automatically entails the correct holographic counterterms, at least in the lower-dimensional examples considered here. This result is reminiscent of the findings by Larsen and McNees \cite{Larsen:2004kf}, who showed for inflationary spacetimes that the requirement of diffeomorphism invariance leads to the correct holographic counterterms at the late time boundary. We intend to apply our procedure to other supersymmetric theories that require holographic renormalization, like pure gravity in AdS$_4$, AdS$_5$ or cosmological topological supergravity in three dimensions. 
One might also apply our program to theories with rigid SUSY on an AdS background, or SUGRA theories with local superconformal invariance. Yet another interesting case are branes in the presence of a Born-Infeld action. Bosonic systems can be treated in the same way by viewing them as truncations of supersymmetric systems. 

Some important open questions are: Why does our program work? Does it work in higher dimensions? Does SUSY require finiteness of response functions like the Brown--York stress tensor? Concerning the last question we recall that the infrared divergences near the AdS boundary are related by duality to the ultraviolet divergences in the boundary theory. SUSY has been quite successful in curing ultraviolet divergences in various theories, and perhaps this is why local SUSY without boundary conditions is capable to produce holographic counterterms.

\section*{Acknowledgments}

We thank Dmitry Belyaev, Simone Giombi, Robert McNees, Leonardo Rastelli and Kostas Skenderis for discussion.

DG was supported by the project MC-OIF 021421 of the European Commission under the Sixth EU Framework Programme for Research and Technological Development (FP6). 
Research at the Massachusetts Institute of Technology is supported in part by funds provided by the U.S. Department of Energy (DoE) under the cooperative research agreement DEFG02-05ER41360.
During the final stage DG was supported by the START project Y435-N16 of the Austrian Science Foundation (FWF).

The research of PvN is supported by the NSF grant PHY-0653342.



\end{document}